\documentclass[12pt]{article}
\usepackage{amsmath}
\usepackage{amssymb}
\usepackage{enumerate}
\begin{document}

\vskip 2cm
\begin{center}
{\bf \Large{Basic Brackets of a 2D Model for the Hodge Theory without its Canonical Conjugate Momenta }}
\vskip 2.8cm

{\sf R. Kumar$^{(a)}$\footnote{Present address: S. N. Bose National Centre for Basic Sciences,
              Block JD, Sector III,\vskip 0.1cm  \hskip 0.3cm Salt Lake, Kolkata$-$700098, India}, 
S. Gupta$^{(a)}$\footnote{Present address: Instituto de F\'\i sica, Universidade de S\~ao Paulo,
              C. Postal 66318, \vskip 0.1cm \hskip 0.3cm 05314-970  S\~ao Paulo, SP, Brazil}, 
R. P. Malik$^{(a,b)}$}\\
$^{(a)}$ {\it Physics Department, Centre of Advanced Studies,}\\
{\it Banaras Hindu University, Varanasi - 221 005, (U.P.), India}\\
$^{(b)}$ {\it DST Centre for Interdisciplinary Mathematical Sciences,}\\
{\it Faculty of Science, Banaras Hindu University, Varanasi - 221 005, India}\\
{\small {\sf {e-mails: rohit.kumar@bose.res.in; saurabh@if.usp.br;  rpmalik1995@gmail.com}}}
\end{center}

\vskip 1.5cm

\noindent
{\bf Abstract:} We deduce the canonical brackets for a two $(1 + 1)$-dimensional (2D) free Abelian
1-form gauge theory by exploiting the beauty and strength of the continuous symmetries of a
Becchi-Rouet-Stora-Tyutin (BRST) invariant Lagrangian density that respects, in totality, 
six continuous symmetries. These symmetries entail upon this model to become a field theoretic example
of Hodge theory. Taken together, these symmetries enforce the existence of exactly the same 
canonical brackets amongst the creation and annihilation operators that are found to exist within the standard
canonical quantization scheme. These creation and annihilation operators
appear in the normal mode expansion of the basic fields of this theory. In other words,
we provide an alternative to the canonical method of quantization for our present model of Hodge
theory where the continuous {\it internal} symmetries play a decisive role. We conjecture that our method of
quantization is valid for a class of field theories that are tractable physical examples for the Hodge theory. 
This statement is true in any arbitrary dimension of spacetime.   \\



\newpage

\section{Introduction}

Symmetry principles, through the ages, have helped physicists to unravel some
of the deepest mysteries of the nature. 
It is well-known, for instance, that the 
symmetries govern interactions [1]. They lead to the conservation 
laws in the realm of classical and quantum systems
and dictate selection rules in the context of the latter. 
In our present investigation, we establish that the continuous symmetries lead 
to the derivation of basic (anti)commutators that are at the 
heart of the covariant canonical quantization of a gauge theory within the 
framework of Becchi-Rouet-Stora-Tyutin (BRST) formalism.\\

The (anti)BRST symmetries and the related geometry (see. e.g. [2-8]), even though 
discovered (and reformulated) nearly four decades ago, are still relevant in the 
context of superstring theories, supersymmetric (SUSY) gauge theories, perturbative
quantum gravity, topological field theories, higher $p$-form ($p = 2, 3...$) gauge 
theories (see, e.g. [9-20]). The deep connections 
of this technique with the ideas of superspace formalism [19,20], its interpretation
in the language of differential geometry; its very useful application in the quantization of
gauge and reparametrization invariant theories, etc., have always kept this formalism at the forefront 
of research in theoretical high energy physics. In our present endeavor, we touch upon a novel
aspect of this technique where it provides an alternative to the mathematical definition of
canonical conjugate momenta for the canonical quantization of a 2D free Abelian gauge  field theory 
which belongs to a class of field theories which are   models for the Hodge theory.\\

In the canonical method of quantization (for a given field theoretic model), a triplet of 
central ideas are exploited. These are the usual spin-statistics relations, definition of the canonical
conjugate momenta and normal ordering. First, using the usual spin-statistics relation, we distinguish
between the bosonic and fermionic field variables. Second, we compute the canonical conjugate momenta
corresponding to the field variables from the Lagrangian density of a given field theoretic system
and define the (graded) Poisson brackets. The latter are upgraded to the (anti)commutators at
the quantum level. Ultimately, in terms of the normal mode expansions of the basic fields and their
corresponding momenta, the above (anti)commutators turn into the (anti)commutators amongst
the creation and annihilation operators and, then, the basic brackets of the theory ensue.
The physical quantities of interest (e.g. Hamiltonian, conserved charges, etc.), expressed
in terms of the creation and annihilation operators are, finally, normal ordered to make
physical sense.   \\

In our present investigation, we shall utilize the virtues of the usual spin-statistic relations
and normal ordering but we shall {\it not} take the help of the definition of canonical 
conjugate momenta in our central goal of obtaining the correct basic brackets amongst
the creation and annihilation  operators of our present field theoretic model for
the Hodge theory (i.e. 2D free Abelian 1-form gauge theory) within the framework
of BRST formalism. Rather, we shall exploit the beauty and strength of the continuous
symmetries (and their generators) to obtain the correct (anti)commutators amongst
the creation and annihilation operators of our present theory which incorporates
fermionic as well as bosonic field  operators. In fact, it is the strength of
{\it all} the six continuous symmetry transformations for this Hodge theory that
entails upon the basic (anti)commutators to emerge in a very natural fashion. \\

In our present paper, we demonstrate that the (anti-)BRST, (anti-)co-BRST, ghost 
and a bosonic symmetry transformations of a free 2D Abelian 1-form gauge theory
(which happens to be a field theoretic model for the Hodge theory [2]) 
imply the existence of the canonical brackets that are required for the covariant canonical
quantization of the above theory within the framework of BRST formalism [3-6]. We emphasize
that {\it all} the above symmetries, taken together, lead to the derivation of the one
and the same canonical brackets [see, equation (17) below] that are also derived by exploiting 
the {\it usual} canonical method from the Lagrangian density (see, Sec. 5 below). Thus, it is
clear that the multi-faceted usefulness of the continuous symmetries enforce
the existence of covariant canonical brackets, too, for {\it a class of field theories} 
that turn out to be the models for the Hodge theory.\\

The prime factors that have propelled us to pursue our present investigation
are as follows. First and foremost, it is very exciting to note that a set of 
continuous symmetries, in some sense, dictate the basic canonical brackets of a given 
class of theories that are models for the Hodge theory. Second, our present work has 
the potential to be generalized to the case of  4D free Abelian 2-form gauge theory
 which is also endowed with a set of six continuous symmetry transformations [21]. 
Third, the derivation of the basic canonical brackets from the symmetry consideration 
(even though algebraically more involved) is more beautiful than the
usual derivation of the same by exploiting the mathematical definition of the canonical
momenta from a given Lagrangian density. Finally, our present work adds yet another 
glittering feather in the already shinning crown of the theoretical versatility of symmetry principles (because
it is the strength of the latter that the mathematical definition of canonical conjugate momenta has
been replaced by the ideas of continuous symmetries and their generators as the Noether conserved charges). \\

The contents of our paper are organized as follows. In Sec. 2, we discuss 
various continuous symmetry properties of the Lagrangian density of the 2D free 
Abelian 1-form gauge theory. Our Sec. 3 is devoted to the derivation of 
conserved charges which, in turn, are expressed in terms of the creation and 
annihilation operators. The derivation of the canonical 
(anti)commutators by exploiting the basic tenets of symmetry principles has been carried out in Sec. 4.
Our Sec. 5 focuses on the derivation of the canonical brackets from the standard canonical method of quantization
applied to the Lagrangian density of the theory. Finally, in Sec. 6, we make some concluding remarks 
and discuss the uniqueness of the basic brackets in our Appendix A.\\

{\it Notations and convention}: We adopt here the convention 
such that the $(1 + 1)$-dimensional (2D) flat Minkowskian metric $\eta_{\mu\nu}$ is endowed with signatures 
$(+1, -1)$ and $ P \cdot Q = \eta_{\mu\nu}\, P^\mu\, Q^\nu = P_0\, Q_0 - P_i \,Q_i $ is 
the dot product between two non-null vectors $P_\mu$ and $Q_\mu$.  Here the  Greek 
indices $\mu, \nu,... = 0, 1$ and Latin indices $i, j,...= 1$. The 2D $F_{\mu\nu}$ 
has only electric field as its non-vanishing component
(i.e. $F_{01} = - \varepsilon^{\mu \nu}\, \partial_\mu A_\nu = E )$. We take 2D 
Levi-Civita  tensor $(\varepsilon_{\mu \nu})$ with the choice 
$\varepsilon_{01} = + 1 = - \varepsilon^{01}$ and it obeys $\varepsilon_{\mu\nu}\, \varepsilon^{\mu\nu} = - 2!,
\;\varepsilon_{\mu\nu}\, \varepsilon^{\nu\lambda} = \delta_\mu^\lambda$. We also 
have the d'Alembertian operator as $\Box = \partial_0^2 - \partial_1^2$.

\section{Lagrangian formalism: continuous symmetries }

We begin with the (anti-)BRST invariant Lagrangian density for a free 2D Abelian 1-form gauge 
theory in the Feynman gauge:
\begin{eqnarray}
{\cal L}_b &=& - \frac {1}{4} F^{\mu \nu} F_{\mu \nu} - \frac {1}{2} \big(\partial \cdot A\big)^2
- i \partial_\mu \bar C\, \partial^\mu C \nonumber\\
&\equiv&  \frac{1}{2} E^2 - \frac {1}{2} \big(\partial \cdot A\big)^2
- i \partial_\mu \bar C \partial^\mu C,
\end{eqnarray} 
where $F_{\mu \nu} = \partial_\mu A_\nu - \partial_\nu A_\mu$ is the curvature 
tensor derived from the 2-form $F^{(2)} = d\, A^{(1)} \equiv \frac{1}{2}\, 
\big(d x^\mu \wedge d x^\nu\big)\, F_{\mu \nu}$. Here $d = d x^\mu\, \partial_\mu $ (with $d^2 = 0$) 
is the exterior derivative 
and connection 1-form $A^{(1)} = d x^\mu \,A_\mu$ defines the vector potential $A_\mu$. 
The gauge-fixing term $\big[- \frac {1}{2}\,\big(\partial \cdot A\big)^2\big]$ owes its origin 
to the co-exterior derivative $\delta = -\, *\, d\, * $ because
$ \delta \, A^{(1)} \equiv -\, * \,d\, *\, A ^{(1)} = + \, \big(\partial \cdot A\big) $ where the
 $(*)$ operation is the Hodge duality (defined on the 2D Minkowski spacetime manifold). The 
fermionic (anti-)ghost fields $(\bar C) C $ $\big($with $ C^2 = \bar C^2 = 0,\; C \,\bar C + \bar C\, C = 0,$
etc.$\big)$ are required for the validity of the unitarity within the framework of BRST formalism.

The Lagrangian density (1) respects the on-shell $\big($i.e. $ \Box\, C = 0,\, \Box \,\bar C = 0\big)$
nilpotent $\big($i.e. $s_{(a)b}^2 = 0 \big)$ (anti-)BRST symmetry transformations $\big(s_{(a)b}\big)$ as
\begin{eqnarray}
& s_b A_\mu = \partial_\mu C, \qquad s_b C = 0, \qquad s_b \bar C = -\, i \,\big(\partial \cdot A\big),
\qquad s_b E = 0,& \nonumber\\
& s_{ab} A_\mu = \partial_\mu \bar C, \qquad s_{ab} \bar C = 0, \qquad
s_{ab} C = +\, i\, \big(\partial\cdot A\big),\qquad s_{ab} E = 0,&
\end{eqnarray}
where the physical (gauge-invariant) electric field (E), owing its origin to
exterior derivative $ d = d x\, ^\mu \partial _\mu $, remains invariant under the
(anti-)BRST symmetry transformations (2). We have the following on-shell 
($ \Box C = 0,\; \Box \bar C = 0$) nilpotent $\big(s_{(a)d}^2 = 0\big)$, continuous and infinitesimal
(anti-)co-BRST symmetry transformations $s_{(a)d}$ (see, e.g. Ref. [22] and references therein) 
\begin{eqnarray}
& s_d A_\mu = -\, \varepsilon_{\mu\nu}\, \partial^\nu \bar C, \quad s_d \bar C = 0, \quad
s_d C = - \,i\, E, \qquad s_d \big(\partial \cdot A\big) = 0,& \nonumber\\
& s_{ad} A_\mu = -\, \varepsilon_{\mu\nu}\, \partial^\nu  C, \quad s_{ad} C = 0, \quad
s_{ad} \bar C = + \,i\, E, \quad s_{ad}\big(\partial \cdot A\big) = 0, &
\end{eqnarray}
that leave the Lagrangian density (1) quasi-invariant [22]. It should be noted here
that the gauge-fixing term $ \big(\partial \cdot A\big) $, owing its origin to the 
co-exterior derivative, remains invariant under the (anti-)co-BRST 
transformations.

A bosonic symmetry $(s_\omega)$ $\big($as the anticommutator $ \big\{ s_b,\, s_d\big \} \equiv
- \big\{s_{ab},\, s_{ad} \big\} = s_\omega\big)$ leads to the following transformations [22]
\begin{eqnarray}
&& s_\omega A_\mu = \partial_\mu E - \varepsilon_{\mu\nu} \,\partial^\nu \big(\partial \cdot A\big),
\qquad s_\omega E =  \Box \big(\partial \cdot A\big), \nonumber\\
&& s_\omega C = 0, \; \qquad s_\omega\bar C = 0, 
\; \qquad s_\omega \big(\partial \cdot A\big) = \Box E,
\end{eqnarray}
under which the Lagrangian density (1) transforms to a total spacetime derivative (see, e.g. Ref. [22]).
Furthermore, we have an infinitesimal ghost symmetry transformation $(s_g)$ in the theory, namely;
\begin{eqnarray}
&& s_g A_\mu = 0, \quad s_g C = + C, \quad s_g \bar C = - \bar C, \quad
s_g E = 0, \quad s_g (\partial \cdot A) = 0,
\end{eqnarray}
which is derived from the  scale transformations $\big(C \rightarrow e^{+\Lambda}\, C,$ $\bar C \rightarrow e^{- \Lambda}\, \bar C,$ 
$A_\mu$ $\rightarrow e^0\,A_\mu\big) $ where the infinitesimal 
scale parameter $\Lambda$ is spacetime independent and, for the sake of brevity, it has
been set equal to {\it one} in the above transformations. It is, thus, 
crystal clear that we have a set of {\it six} continuous
symmetries in the theory. Two of these $\big($i.e. $s_g,\; s_\omega \big)$ are bosonic in nature
and rest of them $\big($i.e. $s_{(a)b},\; s_{(a)d} \big)$ are fermionic 
$\big(s_{(a)b}^2 = 0,\; s_{(a)d}^2 =  0\big)$. 
The latter property is nothing but the nilpotency of order two that is
associated with the (anti-)BRST and (anti-)co-BRST symmetry transformations.

\section{Conserved charges in terms of the creation and annihilation 
operators: normal ordered expressions}

The continuous symmetry transformations, according to Noether's theorem, lead 
to the derivation of the conserved currents. These, in turn, provide us the 
expressions for the conserved charges $\big($i.e. $Q_r = \int dx J^0_r, 
\; r = b, ab, d, ad, g, \omega \big)$. These charges for our present theory are (see, e.g. Ref. [22]).
\begin{eqnarray}
Q_b &=& \int dx  \Big [\partial_0 \big(\partial \cdot A\big) C - \big(\partial \cdot A\big) \dot C \Big ], 
\quad
Q_{ab} = \int dx  \Big [\partial_0 \big(\partial \cdot A\big) \bar C - \big(\partial \cdot A\big) \dot {\bar C} \Big], \nonumber\\
Q_d &=& \int dx \ \Big [E\, \dot {\bar C} - \dot E \,\bar C \Big],   \qquad\qquad \qquad
Q_{ad} = \int dx \; \Big [E\, \dot C - \dot E \,C \Big ], \nonumber\\
Q_\omega &=& \int dx  \Big [\partial_0 \big(\partial \cdot A\big)\, E - \dot E\, \big(\partial \cdot A\big) \Big],  \;\quad
Q_g = i\, \int dx  \Big [C\, \dot {\bar C} + \bar C \,\dot C \Big ],
\end{eqnarray}
where the dot, on a generic field $\Phi$, denotes the time derivative 
$\big[$i.e. $\dot \Phi = (\partial \Phi)/(\partial t) \big]$.
We lay emphasis on the fact that  these conserved charges have been computed from the Noether conserved current
where the definition of canonical momentum plays {\it no} role at all. In fact, it is
the action principle (i.e. $\delta S = 0$) that plays a decisive role in the above 
derivations.

It is evident, from the Lagrangian density (1), that the basic fields of the theory 
satisfy the following Euler-Lagrange equations of motion:
\begin{eqnarray}
\Box A_\mu = 0, \qquad \Box C = 0, \qquad \Box \bar C = 0.
\end{eqnarray}
The normal mode expansions of these fields, in the phase space of our present theory, are
listed below  (see, e.g. Ref. [23])
\begin{eqnarray}
A_\mu (x,t) &=& \int \frac {dk}{\sqrt {2 \pi \,2 k_0}} \, \Big [a_\mu (k) \, e^{+ i k \cdot x} 
 +  a^\dagger_\mu (k) \, e^{- i k \cdot x} \Big ], \nonumber\\
C(x,t) &=& \int \frac {dk}{\sqrt {2 \pi\, 2 k_0}} \, \Big [c(k) \, e^{+ i k \cdot x} 
 +  c^\dagger (k) \, e^{- i k \cdot x} \Big ], \nonumber\\  
\bar C(x,t) &=& \int \frac {dk}{\sqrt {2 \pi\, 2 k_0}} \, \Big [\bar c(k) \, e^{+ i k \cdot x} 
+   \bar c^\dagger (k) \, e^{- i k \cdot x} \Big ],
\end{eqnarray}
where 2-vector $k_\mu = ( k_0, \; k_1 = k)$ is the momentum vector and 
$ a^\dagger_\mu (k), \; c^\dagger (k)$ and $\bar c^\dagger (k)$ are the creation operators
for a photon, a ghost and an anti-ghost quanta, respectively. The non-dagger operators 
$ a_\mu (k), \; c(k)$ and $\bar c(k)$ stand for the corresponding annihilation operators
for a single quantum.

Plugging in these expansions in the expressions for the charges in (6),
we obtain the following
\begin{eqnarray}
Q_b &=& - \int dk \, k^\mu \, \Big [a^\dagger_\mu (k) \, c(k) 
+  c^\dagger (k) \, a_\mu (k) \Big ], \nonumber\\
Q_{ab} &=& - \int dk \, k^\mu \, \Big [a^\dagger_\mu (k) \, \bar c(k) 
+  \bar c^\dagger (k) \, a_\mu (k) \Big ], \nonumber\\
Q_d &=& - \int dk \; \varepsilon^{\mu \nu} \,k_\mu \; \Big [\bar c^\dagger (k) \; a_\nu(k) 
+  a^\dagger_\nu (k) \, \bar c(k) \Big ], \nonumber\\
Q_{ad} &=& - \int dk \, \varepsilon^{\mu \nu}\, k_\mu \, \Big [c^\dagger (k) \, a_\nu(k) 
+  a^\dagger_\nu (k) \, c(k) \Big ], \nonumber\\
Q_g &=& - \int dk \, \Big[ \bar c^\dagger(k) \, c(k) 
+ c^\dagger (k) \, \bar c(k) \Big], \nonumber\\
Q_\omega &=& i\, \int dk \, \varepsilon^{\mu \nu} \, k_\mu k^\rho \, \Big 
[a^\dagger_\rho (k) \, a_\nu(k)  -  a^\dagger_\nu (k) \, a_\rho(k) \Big ] \nonumber\\
&\equiv & i \int dk \, k^2\, \varepsilon^{\mu \nu} \, a^\dagger_\mu (k) \, a_\nu (k),
\end{eqnarray}
where the normal ordering has been taken into account so that all the creation 
operators are kept towards the left. This ordering renders the above charges
physically sensible. In the above, we have taken into account the expression for the
Dirac $\delta$-function as: $\delta (k - k') = (1/2 \pi) \int dx \; e^{\pm (k - k') \cdot x}$ in the 
explicit computations of charges in equation (9). Thus, we have already exploited one of the key requirements
of the quantization scheme, we have proposed to follow.

\section{Canonical brackets: symmetry considerations}

According to the common folklore in quantum field theory, the conserved charges (6)
(that are derived due to the presence of  continuous symmetries in the theory)
generate the continuous symmetry transformations, as 
\begin{eqnarray}
s_r \Phi = \pm \,i \, \big [\Phi, \; Q_r \big ]_\pm, 
\qquad r = b, ab, d, ad, \omega, g, 
\end{eqnarray}
where $\Phi$ is the generic field of the theory and $Q_r$ are the conserved charges
of the theory [cf. (6)]. The $(\pm)$ signs, as a subscript on the square bracket,
correspond to the (anti)commutators for the generic field $\Phi$ being fermionic
(bosonic) in nature. Obviously, we have used here the usual spin-statistics relations
to differentiate between the bosonic and fermionic fields. Quantum mechanically,
this implies the use of suitable brackets [i.e. (anti)commutators] for the quantization
scheme. It should be noted that, in our present 2D theory, there is no concept of ``spin" because
the Pauli-Lubanski vector cannot be defined in 2D. Thus, we have {\it } not used the phrase
``spin-statistics theorem" which is valid in 4D theory. The  $(\pm)$ signs in front of 
the expression on the r.h.s. $\big($i.e. $\pm \, i \,\big [\Phi, \;Q_r \big ]_\pm \big)$ need explanation. 
The pertinent points regarding the choice of a specific sign
(in front of the bracket) are:
\begin{enumerate}[i.]
\item for $s_r = s_b, \, s_{ab}, \, s_d, \, s_{ad}$, only the negative sign would be 
taken into account $\big($i.e. $s_b A_\mu = -\, i \, \big [A_\mu, \, Q_b \big ],$
 $s_b \bar C  = - \,i \, \big\{ \bar C,\, Q_b \big\} $, etc.$\big)$, and
\item for $s_r = s_g, \; s_\omega,$ the negative sign would be taken into account 
for the bosonic field and the positive sign would be chosen for the fermionic 
field  $\big($e.g. $s_g A_\mu = - \,i\, \big [ A_\mu, \, Q_g \big ], \;
s_g C = +\, i \, \big [ C,\, Q_g \big ] , \;
s_g {\bar C} = +\, i \, \big [\bar C, \,Q_g \big ] ,$ etc.$\big)$.
\end{enumerate}

At this juncture, let us take an example $\big($i.e.  $s_b A_\mu = \partial_\mu C \big)$
to make it clear that the symmetry principles dictate the structure of the 
canonical brackets. Mathematically, this symmetry transformation can be expressed 
as [23]
\begin{eqnarray} 
s_b A_\mu = - \,i \, \big [A_\mu, \; Q_b \big ] \; = \;  \partial_\mu C.
\end{eqnarray}
Now taking the normal mode expansions for $A_\mu $ and $C$ [from equation (8)], 
it is clear that we have the following relationships
\begin{eqnarray} 
\big [Q_b, \,a_\mu (k) \big ] = k_\mu \,c(k), \quad 
\big [Q_b,\, a^\dagger_\mu (k) \big ] = -\, k_\mu \,c^\dagger (k).
\end{eqnarray}
Plugging in the expression for $Q_b$ in terms of the creation and annihilation 
operators [cf. (9)], we obtain 
\begin{eqnarray} 
 && \big [a_\mu (k),\; a^\dagger_\nu (k') \big ] = \eta_{\mu \nu} \, \delta (k - k'), \nonumber\\
 && \big [a^\dagger_\mu (k),\, a^\dagger_\nu (k') \big ] = 0, \quad
\big [a^\dagger_\mu (k),\; c(k') \big ] = 0, \quad  \big [a^\dagger_\mu (k), \;c^\dagger (k') \big ] = 0, \nonumber\\
&& [a_\mu (k),\; a_\nu (k^\prime) ] = 0, \quad \big[a_\mu (k),\; c^\dagger (k^\prime)\big] = 0, \quad
 \big[a_\mu (k),\; c (k^\prime)\big ] = 0. 
\end{eqnarray} 
In exactly similar fashion, the following BRST symmetry  transformations 
\begin{eqnarray}
&& s_b C = -\, i \, \big \{ C,\, Q_b \big\} = 0 \Longrightarrow 
 \big\{Q_b,\, c(k) \big\} = 0, \quad  \big\{Q_b, \, c^\dagger (k) \big\} = 0, \nonumber\\  
&& s_b \bar C = - \,i \, \big \{ \bar C,\, Q_b \big\} = -\, i\, \big(\partial \cdot A\big) \Longrightarrow \nonumber\\
&&\big\{ Q_b, \,\bar c(k) \big\} = \,+\, i\, k^\mu \,a_\mu (k), \qquad 
\big\{ Q_b, \,\bar c^\dagger (k)\big \} = \,-\, i\, k^\mu \,a^\dagger_\mu (k), 
\end{eqnarray}
lead to the derivation of the following brackets
\begin{eqnarray} 
&& \big\{ c^\dagger (k),\; \bar c (k') \big\} = - \,i \, \delta (k - k'), \qquad 
\big\{ c (k), \;\bar c (k') \big\} = 0,  \nonumber\\
&& \big\{ c (k), \;\bar c^\dagger (k') \big\} = +\, i \, \delta (k - k'), \qquad 
\big\{ c^\dagger (k),\; \bar c^\dagger (k') \big\} = 0, \nonumber\\
&& \big [a^\dagger_\mu (k),\; \bar c(k') \big ] = 0,   \quad
\big [a_\mu (k), \;\bar c(k') \big ] = 0,  \quad
\big [a^\dagger_\mu (k), \;\bar c^\dagger (k') \big ] = 0,  \nonumber\\
&& \big [a_\mu (k),\; \bar c^\dagger (k') \big ] = 0, \quad
\big[a_\mu^\dagger (k), \;c(k^\prime) \big] = 0, \quad 
\big[a_\mu (k), \;c (k^\prime)\big] = 0.
\end{eqnarray}
It is worthwhile to point out that the following statements are true, namely; 
\begin{enumerate}[i.]
\item the above brackets have been derived by taking into account (see, e.g. Ref. [23] for details) {\it only} the 
on-shell nilpotent BRST symmetry 
transformations $\big[$i.e. $s_b A_\mu = \partial_\mu C, \; s_b C = 0, \; 
s_b \bar C = -\, i\, (\partial \cdot A) \big]$, and
\item the expressions on the r.h.s. of equations (12) and (14) enforce,
in a definite manner, the choice of
the (anti)commutators in (13) and (15) when we use the expression for $Q_b$ from (9).
\end{enumerate}

We would like to emphasize that the above exercise can be performed with all the 
six conserved charges listed in (6). The relevant (anti)commutators, emerging out from
this algebraic exercise, are as follows
\begin{eqnarray} 
&& \big [Q_{ab}, \;a_\mu (k) \big ] = +\, k_\mu \,\bar c(k), \hskip 1.7cm \big\{Q_{ab},\;\bar c(k) \big\} = 0,\nonumber\\
&& \big [Q_{ab}, \;a^\dagger_\mu (k) \big ] = -\, k_\mu\, \bar c^\dagger (k), \hskip 1.5cm  \big\{Q_{ab},\; \bar c^\dagger (k)\big \} = 0, \nonumber\\
&&  \big\{ Q_{ab},\; c(k)\big \} = -\, i\, k^\mu \,a_\mu (k), \hskip 1.3cm \big\{Q_d, \;\bar c(k) \big\} = 0, \nonumber\\
&& \big\{ Q_{ab}, \;c^\dagger (k) \big\} = +\, i\, k^\mu \,a^\dagger_\mu (k), \hskip 1.2cm  \big\{Q_d,\; \bar c^\dagger (k) \big\} = 0, \nonumber\\
&&  \big [Q_d,\; a_\mu (k) \big ] = -\, \varepsilon_{\mu\nu}\, k^\nu\, \bar c(k), \hskip 1.2cm \big\{Q_{ad}, \;c(k) \big\} = 0,\nonumber\\
&& \big [Q_d,\; a^\dagger_\mu (k) \big ] =  +\, \varepsilon_{\mu\nu}\, k^\nu \,\bar c^\dagger(k),  
\hskip 1.1cm  \big\{Q_{ad}, \;c^\dagger (k) \big\} = 0, \nonumber\\ 
&& \big\{ Q_d, \;c(k) \big\} = -\, i\, \varepsilon^{\mu\nu}\, k_\mu \,a_\nu (k), \hskip 0.9cm \big [Q_g, \;a_\mu (k) \big ] = 0,\nonumber\\
&& \big\{ Q_d,\; c^\dagger (k) \big\} = +\, i\, \varepsilon^{\mu\nu}\, k_\mu \,a^\dagger_\nu (k),
\hskip 0.8cm \big [Q_g, \;a^\dagger_\mu (k) \big ] = 0, \nonumber\\
&&  \big [Q_{ad},\, a_\mu (k) \big ] = - \varepsilon_{\mu\nu}  k^\nu \,c(k), 
\hskip 1.3cm \big[ Q_g,\, \bar c^\dagger(k) \big ] = - i \, \bar c^\dagger (k),  \nonumber\\
&& \big [Q_{ad}, \;a^\dagger_\mu (k) \big ] =  +\, \varepsilon_{\mu\nu} \,k^\nu\, c^\dagger(k),
\hskip 1.cm \big[ Q_g,\, \bar c(k) \big ] = - i \,\bar c(k) \nonumber\\ 
&&  \big\{ Q_{ad},\; \bar c(k)\big \} =  +\, i\, \varepsilon^{\mu\nu} \,k_\mu \,a_\nu (k), 
\hskip 0.8cm \big[Q_\omega, \;\bar c(k) \big ] = 0, \nonumber\\
&& \big\{ Q_{ad},\; \bar c^\dagger (k)\big \} = -\, i\, \varepsilon^{\mu\nu}\, k_\mu \,a^\dagger_\nu (k),
\,\hskip 0.6cm \big[ Q_\omega, \;\bar c^\dagger (k) \big ]  =  0, \nonumber\\
&&  \big[ Q_g, \,c(k) \big ] = +  i c(k), \,\hskip 2.6cm
\big[ Q_g, \,c^\dagger(k) \big ] = +  i c^\dagger (k), \nonumber\\
&& \big[Q_\omega, a^\dagger_\mu (k) \big ] = - i k^2\, \varepsilon_{\mu\nu} (a^\nu)^\dagger (k), 
\,\hskip 0.7cm \big[Q_\omega, \;c(k) \big ]  = 0, \nonumber\\
&& \big[Q_\omega, a_\mu (k) \big ] = - i k^2\, \varepsilon_{\mu\nu} a^\nu (k), \hskip 1.2cm  \big[Q_{\omega}, \;c^\dagger (k) \big] = 0.
\end{eqnarray}
The outcome of all the above (anti)commutators, with the help of the
normal mode expansions (8) and the expressions for the charges in (9), 
lead to the following non-vanishing basic brackets
\begin{eqnarray}
&& \big [a_\mu (k),\; a^\dagger_\nu (k') \big ] = \eta_{\mu \nu} \, \delta (k - k'), \nonumber\\
&& \big\{ c (k), \;\bar c^\dagger (k') \big\} = +\, i \, \delta (k - k'), \nonumber\\
 && \big\{ c^\dagger (k), \;\bar c (k') \big\} = - \,i \, \delta (k - k'),  
\end{eqnarray} 
within the framework of BRST formalism. All the rest of the (anti)commutators 
turn out to be zero. To summarize,  we have already utilized {\it all} the ingredients
of {\it our} quantization scheme (without any use of the mathematical definition of canonically conjugate momenta
{\it anywhere} in our discussions).

\section{Standard canonical method: Lagrangian formalism }

It is evident that the canonical conjugate momenta from the Lagrangian density (1),
for the basic fields of the theory, are
\begin{eqnarray}
&& \Pi^\mu = \frac {\partial {\cal L}_{(b)}}{\partial (\partial_0 A_\mu )} = -\, F^{0 \mu}
- \eta^{0 \mu} \big(\partial \cdot A\big),\nonumber\\
&& \Pi_C = \frac {\partial {\cal L}_{(b)}}{\partial (\partial_0 C)} = +\, i\, \dot {\bar C},
\qquad \Pi_{\bar C} = \frac {\partial {\cal L}_{(b)}}{\partial (\partial_0 \bar C)} 
= - \,i \,\dot C.
\end{eqnarray}
As a result, we have the following canonical brackets: 
\begin{eqnarray}
&& \big[A_\mu (x,t), \;\Pi_\nu (x',t)\big] = i\, \eta_{\mu\nu}\, \delta (x - x'), \nonumber\\
&& \big\{\bar C(x,t),\; \Pi_{\bar C} (x', t) \big\} = i\, \delta (x - x') \Longrightarrow 
\big\{\bar C(x,t), \;\dot{C} (x', t) \big\} = - \,\delta (x - x'), \nonumber\\
&& \big\{C(x,t),\; \Pi_C (x', t)\big \} = i\, \delta (x - x') \Longrightarrow 
\big\{ C(x,t), \;\dot{\bar C} (x', t)\big \} =  \delta (x - x'). 
\end{eqnarray} 
All the rest of the brackets are zero. The top entry, in the above, implies the following 
commutators in terms of the components of the 2D gauge field $A_\mu$ and the corresponding
conjugate momenta, namely; 
\begin{eqnarray}
\big[A_0 (x, t),\, \big(\partial \cdot A\big) (x^\prime, t) \big] &=& - \,i\, \delta (x - x^\prime), \nonumber\\
\big[A_i (x, t),\; E_j (x^\prime, t) \big] &=&  i\, \delta_{ij} \,\delta (x - x^\prime).\quad
\end{eqnarray}
The above form of commutators would be useful later.

To simplify, the rest of our computations, we re-express the normal mode expansions
of the basic fields [cf. (8)], as  [24]
\begin{eqnarray}
A_\mu (x, t) &=& \int dk \; \Big[ a_\mu (k) \; f^* (k, x) \; 
+ \; a^\dagger_\mu (k) \; f(k, x) \Big], \nonumber\\
C (x, t) &=& \int dk \; \Big[ c (k) \; f^*(k, x) \; + \;  c ^\dagger (k) \;  f(k, x)
\Big], \nonumber\\
\bar C (x, t) &=& \int dk \; \Big[ \bar c (k) \; f^*(k, x) \; 
+ \; \bar c ^\dagger (k) \; f(k, x) \Big], 
\end{eqnarray}
where the new functions:
\begin{eqnarray}
f(k, x) = \frac {e^ {-i k \cdot x}} {\sqrt {2 \pi \,2 k_0}}, \qquad
f^*(k, x) = \frac {e^ {i k \cdot x}} {\sqrt {2 \pi \,2 k_0}},
\end{eqnarray}
form an orthonormal set because they satisfy [24]
\begin{eqnarray}
&& \int dx \;  f^*(k, x) \, i\, \overleftrightarrow{\partial_0} \; f(k', x) = \delta (k - k'), \;\quad
\int dx \; f^*(k, x) \, i \,\overleftrightarrow{\partial_0} \; f^*(k', x) = 0, \nonumber\\
&& \int dx \; f(k, x) \, i\, \overleftrightarrow{\partial_0} \; f(k', x) = 0,
\end{eqnarray} 
where we have taken into account the following definition 
\begin{eqnarray}
A \;  \overleftrightarrow{\partial_0} \; B =  A \,\big(\partial_0 B\big) -  \big(\partial_0 A\big)\, B, 
\end{eqnarray}
for the operator $\overleftrightarrow{\partial_0}$ between two non-zero variables $A$ and $B$. 
Using the above relations, it is straightforward to check that 
\begin{eqnarray}
a_\mu (k) &=& \int dx \; A_\mu (x, t) \; i\, \overleftrightarrow{\partial_0} \; f(k, x), \quad
a^\dagger_\mu (k)  = \int dx \;  f^*(k, x) \, i\, \overleftrightarrow{\partial_0} \; A_\mu (x, t), \nonumber\\
c^\dagger (k) &=& \int dx \; f^*(k, x) \, i\, \overleftrightarrow{\partial_0} \; C (x, t), \quad \;
\bar c^\dagger (k) = \int dx \; f^*(k, x) \, i\, \overleftrightarrow{\partial_0} \; \bar C (x, t), \nonumber\\
c(k) &=& \int dx \; C(x, t) \, i\, \overleftrightarrow{\partial_0} \; f(k, x), \quad \quad
\bar c(k) = \int dx \; \bar C(x, t) \, i\, \overleftrightarrow{\partial_0} \; f(k, x).
\end{eqnarray}
Thus, we have expressed the creation and annihilation operators in terms of the fields
and the orthonormal functions $f(k, x)$ and $f^*(k, x)$.

At this stage, a few comments are in order. First and foremost, it is straightforward to
check that only the canonical brackets (17) survive in the explicit computation. Second,
there exist {\it six} anticommutators from the four fermionic operators $ c(k),\; c^\dagger (k),\;
\bar c(k),\; \bar c^\dagger (k)$. Out of which, four would be zero because of the 
orthonormality relations (23) and because of the fact that $ C^2 = \bar C^2  = 0, \;
\big\{ C (x, t), \;\dot C (x^\prime, t)\big \} = 0, \;
\big\{ \bar C (x, t), \;\dot {\bar C} (x^\prime, t) \big\} = 0$. Third, there exist {\it three} basic
commutators from $ a_\mu (k)$ and $a_\mu^\dagger (k)$. Out of which, two would turn out
to be zero because the commutation relations in (20) can be recast in the form
$\big[ A_\mu (x, t),\; \dot A_\nu (x^\prime, t)\big] = - \,i\, \eta_{\mu\nu}\, \delta (x - x^\prime)$
due to the fact that (i) $\dot A_0 = \big(\partial \cdot A\big) \,+\, \partial_i\, A_i$  and
$\dot A_i = E_i \, +\, \partial_i\, A_0$, and (ii) the spatial derivative of the gauge
field $A_\mu$ commutes with itself.

It is straightforward to check that the canonical brackets of (19) and (20)
[that are derived from the Lagrangian density (1)] lead to the derivation of the same brackets
that are listed in (17). Thus, we conclude that the basic canonical brackets [cf. (17)]
between the creation and annihilation operators of the bosonic and fermionic fields of the 
theory can be derived from (i) the continuous symmetry considerations, and (ii) by exploiting 
the definition of momenta from the Lagrangian density of the theory.

\section{Conclusions}

In our present investigation, the key ideas that have been exploited for the quantization
scheme are the usual spin-statistics relations, normal ordering (in the expressions for the conserved
charges) and the key concepts  of the continuous symmetry transformations (and  corresponding
generators). The last ingredient of the above quantization scheme is the {\it novel} one and it differs
from the standard method of  canonical quantization scheme where the (graded) Poisson brackets
(defined with the help of definition of the conjugate momenta) are promoted to the (anti)commutators in 
addition to the helps coming out from the usual spin-statistics relations and normal ordering.
It is worthwhile to mention that, in a 2D Minkowski space, there is {\it no} concept of spin 
because the Pauli-Lubanski vector cannot be defined on a 2D spacetime manifold. Thus,
the implication of the usual spin-statistics relation  is the existence of (anti)commutators at the quantum level
in our present 2D field theoretic model of Hodge theory. \\

One of the most beautiful observations in our present endeavor is the emergence of  one and the
same set of non-vanishing basic canonical brackets [cf. (17)] from {\it all} the continuous symmetry
transformations present in the theory. These basic canonical brackets are found to be 
{\it unique} and any other alternatives/deformations to them would {\it not} work with {\it all} 
the continuous symmetries. Even though the continuous 
symmetry transformations (and corresponding generators) look completely different,
the hidden basic brackets (17) [that emerge from the application of (10)] are exactly the same.\\

The above key observation ensures that the continuous symmetry transformations of a field theoretic
model for the Hodge theory encode in their folds the basic (anti)commutators corresponding to the fermionic
and bosonic fields of this theory. To the best of our knowledge, our method of derivation of 
the basic brackets in (17) is a {\it novel} observation in the realm of the quantization scheme of a gauge
theory. It should be noted that two of us (SG and RK)  have already generalized our present method
of quantization to 4D free Abelian 2-form gauge theory [21] 
(which also happens to be a field theoretic model for Hodge theory [25]) and
we have also obtained the exact basic (anti)commutators in the context of quantization scheme for 
a 2D interacting Abelian  1-form $U(1)$ theory where the Dirac fields are also present [26].
The latter fields couple with the $U(1)$ gauge field in a gauge invariant manner. In a very recent publication
[27], we have derived the basic brackets at the level of creation and annihilation operators in the 
case of a toy model for the rigid rotor which is {\it also} is an example of the Hodge theory.  \\

Our method of quantization is completely different from the derivation of canonical 
brackets by considering the equations of motion for the quantum mechanical system 
of harmonic oscillator by Wigner [28]. In our approach, we do not use the equations 
of motion anywhere except in the proof of nilpotency of the (anti-)BRST and (anti-)dual 
BRST symmetries [cf. (2) and (3)] respectively. Rather, we exploit the ideas of symmetry 
principle and the definition of generators in our approach. In fact, it is the existence 
of {\it six} continuous symmetries of our present model for the Hodge theory that entails upon the emergence 
of basic canonical  brackets (17) {\it uniquely} at the level of creation/annihilation operators
(that appear in the normal mode expansions of the fields). It would be very nice endeavor
for us to apply our method to the quantization of other challenging filed theoretic  models of Hodge theory 
in higher  dimensions of spacetime. We are intensively involved, at the moment, with such kinds of problems
 and our results would be reported elsewhere [29].  \\

\vskip 1.5cm

\noindent
{\bf Acknowledgements} \\

\noindent
SG and RK would like to gratefully acknowledge the financial
support from CSIR and UGC, New Delhi, Government of India, respectively.\\

\vskip 1.5cm

\section*{Appendix A: On the uniqueness of the basic brackets}
Here we describe very briefly  the uniqueness  of the canonical brackets 
(taken as the (anti)commutation relations) amongst the creation and annihilation 
operators which have been derived in the equation (17). Exploiting the general 
definition (10) of the generator and the transformations (2) and (3), it can be checked that the following 
alternative (to (17)) (anti)commutators (with $A(k) = A^\dagger (k)$ as a parameter)    
\begin{eqnarray}
\big [a_{\mu}(k), a^{\dagger}_{\nu}(k')\big] &=& \eta_{\mu\nu} \; \delta(k-k') 
+ A(k) \; a^{\dagger}_{\nu}(k)\; a_{\mu}(k) \; \delta(k-k'), \nonumber\\
\{c^\dagger (k), \bar c(k') \} &=& - \, i \, \delta(k - k') + A(k) \;c^\dagger (k)\; \bar c(k) \;\delta (k - k'),\nonumber\\  
\{{\bar c}^\dagger (k), c(k') \} &=& i \; \delta(k - k') + A(k) \;{\bar c}^\dagger (k)\; c(k) \;\delta (k - k'),\nonumber\\ 
\big[c(k), a_{\mu}(k')\big] &=& A(k) \; a_{\mu}(k) \; c(k) \; \delta(k-k'), \nonumber\\
\big [c^{\dagger}(k), a^{\dagger}_{\mu}(k')\big] &=&
- A(k) \; c^{\dagger}(k) \; a^{\dagger}_{\mu}(k) \; \delta(k-k'), \nonumber\\
\big[{\bar c}^\dagger (k), a^\dagger_\mu (k')\big] &=& 
- A(k)\;{\bar c}^\dagger (k)\;a^\dagger_\mu (k)\; \delta (k - k'),\nonumber\\ 
\big[\bar c (k), a_\mu (k')\big]  &=&  A(k)\; a_\mu (k)\;\bar c(k) \;\delta (k - k'),\nonumber\\
\{{\bar c}^\dagger (k), \bar c(k') \} &=&  A(k) \;{\bar c}^\dagger (k)\;  \bar c(k) \;\delta (k - k'),\nonumber\\ 
\{c^\dagger (k), c(k') \} &=& A(k) \; c^\dagger (k)\; c(k) \;\delta (k - k'),\nonumber\\ 
\{c^{\dagger}(k), c(k')\} &=& A(k) \;c^{\dagger}(k) \;  c(k) \; \delta(k-k'),\nonumber\\ 
\big[a_{\mu}(k), a_{\nu}(k')\big ] &=& 0, \qquad \big[a^{\dagger}_{\mu}(k), a^{\dagger}_{\nu}(k')\big] = 0, \nonumber\\
\big[c^{\dagger}(k), a_{\mu}(k')\big] &=& 0, \qquad \big [c(k), a^{\dagger}_{\mu}(k')\big] = 0,\nonumber\\
\big[\bar c(k), a^\dagger_\mu (k')\big] &=& 0,\qquad \big[{\bar c}^\dagger (k), a_\mu (k')\big] = 0,\nonumber\\
\{\bar c(k), \bar c(k') \} &=& 0,\qquad \{{\bar c}^\dagger (k), {\bar c}^\dagger (k') \} = 0,\nonumber\\
\{\bar c(k), c(k') \} &=& 0,\qquad \{{\bar c}^\dagger (k), c^\dagger (k') \} = 0,\nonumber\\
\{c(k), c(k') \} &=& 0,\qquad \{c^\dagger (k), c^\dagger (k') \} = 0,
\end{eqnarray} 
are consistent
with the appropriate brackets listed in (12), (14) and (16) with the nilpotent (anti-)BRST
and (anti-)co-BRST charges when we exploit the analogue of (11) in a suitable fashion. In other words, as far 
as the nilpotent (anti-)BRST and (anti-)co-BRST symmetries are concerned, the brackets (26) are as good 
as the canonical brackets derived in (17).  However, the brackets (26) fail miserably to 
be consistent with the other two conserved charges $Q_\omega$ and $Q_g$ [cf. (9)] when
we exploit the equations (4), (5), (10) in the analogue of (11) for the test of consistency as well as 
preciseness of the brackets (26) in the derivation of the appropriate brackets 
[i.e. (anti)commutators] of (16).

To corroborate the above assertions,  let us take a couple of explicit examples with 
$Q_\omega$ and $Q_g$ (which generate the  bosonic and ghost transformations). 
For instance, the following two brackets from (16)
\begin{eqnarray}
\big [Q_\omega, \, a_\mu (k) \big ] = - i k^2 \varepsilon_{\mu\nu} \, a^\nu(k), \qquad
\big[Q_g, \, c(k)\big] = i c(k),
\end{eqnarray}
must be satisfied by the alternative brackets given in (26). The l.h.s. of the first 
commutator \big(i.e. $ \big [Q_\omega, \; a_\mu (k) \big ]$ $= 
- i k^2 \; \varepsilon_{\mu\nu} \; a^\nu(k)$ \big) of the above equation (27) implies the
following commutator:
\begin{eqnarray} 
\big [Q_\omega, \; a_\mu (k) \big ]  &=&  i \int  dk' \; k'^2 \varepsilon^{\rho\nu} 
\Bigl( a_\rho^\dagger (k') \;  [a_\nu (k'), \;  a_\mu (k) ]\nonumber\\
 &+&  [a_\rho^\dagger (k'), \; a_\mu (k) ] \; a_\nu (k') \Bigr). 
\end{eqnarray}
Using the appropriate brackets from (26), we obtain 
\begin{eqnarray} 
 \big [Q_\omega, \; a_\mu (k) \big ] &=& - i  k^2 \varepsilon_{\mu\nu} \; a^\nu (k) 
- i  k^2 \; \varepsilon_{\rho\nu}\; A(k) \;(a^\rho)^\dagger (k)\; a_\mu (k) \; a^\nu (k).
\end{eqnarray}
It is obvious that the r.h.s. of the above equation does not match with the required result
of (27). Thus, the brackets (26) are not consistent. They can be consistent if and only if 
$A(k) = 0$ which, ultimately, implies the uniqueness of  non-vanishing basic canonical brackets (17).

To check the sanctity and preciseness of the brackets (26),
now let us take the second commutator \big (i.e. $\big[Q_g, c(k)\big] = i\; c(k)$ \big) of
the above equation (27). The l.h.s. of the commutator is as follows 
\begin{eqnarray}
 \big[Q_g, c(k)\big] &=& - \int dk'\Big(- \{{\bar c}^\dagger (k'), c(k)\}\; c(k') 
+ {\bar c}^\dagger (k')\;\{c(k'), c(k)\} \nonumber\\
&-& \{c^\dagger (k'), c (k)\} \;\bar c (k') + c^\dagger (k') \;\{ \bar c(k'), c(k)\}\Big).
\end{eqnarray}
In the above equation, we express the ghost charge in terms of the creation and annihilation operators 
[cf, (9)] and choose the appropriate brackets from (26). It is evident that 
$\{c(k), c(k')\} = 0, \{\bar c(k), c(k')\} = 0 $ [cf. (26)]. 
With these inputs, it can be checked that 
\begin{eqnarray}
\big[Q_g, c(k)\big] &=& i c(k) + c^\dagger (k) \;A(k)\; c(k)\;\bar c(k).
\end{eqnarray}
The above equation is in conflict  with the requirement of the r.h.s. of equation (27) 
[unless $A(k) = 0$]. Thus, the (anti)commutators in 
(26) are {\it not} consistent in proving equation (27) in its exact form.

Ultimately, we conclude that, even though the 
brackets (26) are consistent with all the requirements connected with the charges $Q_{(a)b}$
and $Q_{(a)d}$, they are {\it not} consistent with charges $Q_\omega$ and $Q_g$. 
We wish to lay emphasis  on the fact that we have taken {\it only}
two brackets in equation (27). However, as it turns out, all the rest of the relevant 
brackets with $Q_\omega$ and $Q_g$ are found to be inconsistent. Hence
the canonical (anti)commutators of equation (17) are {\it unique} in the sense that they are
consistent with all the {\it six} continuous symmetries and corresponding conserved charges.
In general, there might exist many kinds of deformations like (26). However, the consistency with all
the {\it six} continuous symmetries would always lead to the derivation of (17).

\end{document}